\documentclass[a4paper,11pt]{article}
\usepackage{amsmath,graphicx,color,mathrsfs,subfigure,multicol}

\def\circa#1{\,\raise.3ex\hbox{$#1$\kern-.75em\lower1ex\hbox{$\sim$}}\,}
\makeatletter

\usepackage{epsfig} 
\usepackage{cite}
\usepackage{amsmath,amsfonts,amssymb}

\def\endignore{}
\def\ignore #1\endignore{} 

\def \gsim{\mathrel{\vcenter
     {\hbox{$>$}\nointerlineskip\hbox{$\sim$}}}}

\newcommand{\Mp}{M_P}

\newcommand{\vp}{\varphi}

\newcommand{\fnlf}{ f^{(4)}_{\rm NL} }

\newcommand{\fnl}{f_{\rm NL}}

\newcommand{\calP}{{\cal P}}

\def\bea{\begin{eqnarray}}
\def\eea{\end{eqnarray}}

\def\bkone{{\bf k_1}} 
\def\bktwo{{\bf k_2}}

\def\picube{(2\pi)^3}
\newcommand{\sdelta}[1]{\!\delta^{\,3}(\mathbf{#1})}

\def\be{\begin{equation}}
\def\ee{\end{equation}}
\def\bea{\begin{eqnarray}}
\def\eea{\end{eqnarray}}

\def\endignore{}
\def\ignore #1\endignore{} 

\def\bd{\begin{displaymath}}
\def\ed{\end{diplaymath}}

\def\({\left(}
\def\){\right)}

\def\bea{\begin{eqnarray}}
\def\eea{\end{eqnarray}}
\def\be{\begin{equation}}
\def\ee{\end{equation}}

\def\bd{\boxdot}

\begin{document}

\hfill HD-THEP-09-8
\begin{center}
{\LARGE \bf
Non-Gaussianity beyond slow roll in multi-field inflation 
}\\[1cm]

{
{\large\bf Christian T.~Byrnes~\footnote{{ E-mail}:  
\texttt{c.byrnes@thphys.uni-heidelberg.de}},\, Gianmassimo Tasinato~\footnote{{ E-mail}: 
\texttt{g.tasinato@thphys.uni-heidelberg.de}
}
}
}
\\[7mm]
{\it Institut f\"ur Theoretische Physik, 
 Philosophenweg 16 and 19, 
69120 Heidelberg, Germany}
\vspace{-0.3cm}

\vspace{1cm}


\end{center}
\begin{quote}

{
We study the non-Gaussianity generated during multiple-field inflation. We provide an exact
expression for the bispectrum parameter $f_{NL}$ which is valid beyond the slow-roll regime, valid
for certain classes of inflationary models. We then study a new, exact multi-field inflationary
model
considering a case where the bispectrum grows to observable values at the end of inflation. We show
that in this case the trispectrum is also large and may even provide the dominant signal of
non-Gaussianity.

}

\end{quote}

\section{Introduction}


The inflationary paradigm predicts
an almost scale--invariant spectrum of nearly Gaussian
perturbations,
in excellent agreement with
 observations \cite{wmap5}. On the other hand,
many different models of inflation  fit  present day
data very well, and further information is needed in order
to test and exclude most of them.  Non-Gaussianity, a signal of
 the non-linearity of the 
perturbations, has emerged in recent
years 
as an important discriminator between various inflationary models
\cite{Komatsu:2009kd,Bartolo:2004if}. 

\smallskip

There are various known
 ways to generate  large non-Gaussianity within the
 inflationary
 paradigm. Models of inflation with non-canonical kinetic terms  are  characterized  by  an equilateral shape of non-Gaussianity,
   generated on sub-Hubble scales \cite{DBI}.
  Non-local inflation can also lead to large non-Gaussianity
  with local shape \cite{cline}. 
   These examples
are non-generic, 
 since inflaton perturbations 
  are usually close to Gaussian at Hubble exit
\cite{Maldacena:2002vr,Seery:2005gb}. 
 On the other hand,  a local
shape of non-Gaussianity can be  generated 
during the evolution of fluctuations
on super-horizon scales~\footnote{
Different possible shape dependences of the bispectrum
(3--point function) are considered in \cite{Fergusson:2008ra}.}.
  Models have
been proposed to produce a large non-Gaussianity
with this shape 
{\it after} inflation, such as inflationary models
with an inhomogeneous end
 \cite{Bernardeau:2002jy,endofinflation},
modulated preheating or reheating \cite{modulatedreheating}, and the curvaton scenario
\cite{curvaton}. However all of these methods require the existence of additional, very light
scalar fields, which play no role during inflation but have a significant effect on generating the
primordial curvature perturbation after inflation ends. 
Arguably,
models which generate  large, {\it primordial}
non-Gaussianity during inflation
are therefore more economical.

\smallskip
Multi-field models of inflation, in which the curvature perturbation
evolves at superhorizon scales, are able, at least in principle,
 to generate large non-Gaussian fluctuations
 during inflation. Most of the
 previous analyses of primordial
  non-Gaussianity 
  in the multi-field case focused on
 separable Ans\"atze for the scalar field
potential, either sum separable (for two-fields see \cite{VW}, for an arbitrary number of fields see
\cite{Battefeld:2006sz,Seery:2006js}) or product separable \cite{Choi:2007su},
and rely  on the slow-roll approximation. Examples in which
 $f_{NL}$ can grow large during slow--roll inflation for these classes of potentials were given
\cite{Byrnes:2008wi}, as well as the general requirements
 that the initial conditions of the
background inflationary trajectory must satisfy. Numerical
analyses have also been made,  going
 beyond the
assumption of separability and/or slow roll \cite{Rigopoulos:2005us,Yokoyama:2007uu}, but no example
of a model giving rise to a large non-Gaussianity at the end 
of inflation has been found in
those cases.

\smallskip

In this paper, we consider models of inflation driven 
by multiple scalar fields with canonical
kinetic terms. We study  the properties of  
 non-Gaussianity, parameterised 
  by the quantity  
  $f_{NL}$ characterising
 the bispectrum, 
  generated 
 by the evolution of scalar perturbations on super Hubble scales.  
 We develop a new method 
 that provides, for certain classes of models, 
 analytical 
 expressions
for $f_{NL}$  valid in regimes
   {\it beyond} a slow--roll approximation.
    Our method is based on the first--order Hamilton--Jacobi formalism developed by Salopek and Bond \cite{SB},
which allows us  to express inflationary observables in the
multi-field case, without having to  focus on a  slow--roll
regime (see \cite{kinney} for
a similar
 application of this formalism to the single-field case). 
    
   The main idea on which our arguments 
   are based is     to focus on  inflationary multi-field trajectories
  in  which the {\it Hubble rate}, and not the potential,
   is separable. 
  This approach allows us to analytically
   study cases where the 
  non-Gaussian parameter
  $f_{NL}$ becomes large,  in regimes in which the slow--roll
parameters are enhanced after Hubble exit, but before inflation ends.
In particular, 
we provide analytical 
equations that express the non-linearity parameter
$\fnl$, in terms of quantities that generalise the slow-roll
parameters during inflation.  
By means of the same method, we are also able to find
new, exact multi-field  homogeneous 
solutions,  that satisfy our Ansatz, for which a large amount of  non-Gaussianity can be generated
before the end of inflation.  This allows us to apply the general 
formulae we find on a concrete example, that shows how
the $f_{NL}$ parameter, and also both non-linearity parameters which characterise the
trispectrum, can grow parametrically
large as a function of quantities that characterise
the model. In some cases the trispectrum through $g_{NL}$ can give the dominant signal of
non-Gaussianity.

  \smallskip

The plan of this paper is as follows: first we briefly recap the $\delta N$ formalism, which allows
us to compute the evolution of perturbations on large scales. In section \ref{themethod} we present
a method to calculate $f_{NL}$ exactly for some models, by using the Hamilton-Jacobi equations. In
section \ref{exactsolutions} we consider some explicit multi-field models of inflation for which we
find exact solutions, and in one case we show that the non-Gaussianity can become observably large
by the end
of inflation. Finally we conclude in section \ref{conclusions}.

\section{Brief recap of $\delta N$ approach to evolution of perturbations}\label{briefreview}

In this paper 
we use the so-called  $\delta N$ approach
to study
the evolution of curvature perturbations. This formalism
has been found to be 
convenient  for studying non-Gaussianity
produced during inflation, and has been implemented
in several previous studies. In this section we give
a very brief review of the formulae necessary for 
the following discussion.

The starting point is the expression for the
 number of e-foldings of expansion, given by 
\be N=\int^{t_{{c}}}_{t_*} H(t)dt,
\ee
evaluated from an initial spatially flat hypersurface to a final
uniform density hypersurface. The
perturbation in the number of $e$-foldings, $\delta N$, is the difference between the curvature
perturbations on the initial and final hypersurfaces. 
We take the initial time to be
Hubble exit during inflation, denoted by $t_*$. 
The final time $t_c$ corresponds
to a uniform density hypersurface just before the end of inflation.
The curvature perturbation is given by \cite{Sasaki:1998ug,LR}
%
\be\label{deltaN} 
\zeta=\delta N= \sum_I
N_{,I}\delta\vp_{I*}+\frac12\sum_{IJ}N_{,IJ}\delta\vp_{I*}\delta\vp_{J*}+\cdots\,, \ee
where $N,_I=\partial N/(\partial \vp^I_*)$ and the index $I$ runs over all of the fields. The
power spectrum and bispectrum are defined (in Fourier space) by \footnote{Note that different sign
conventions of $f_{NL}$ have been used. We use the same sign convention as WMAP \cite{wmap5}, which
is opposite to that used in \cite{VW,Choi:2007su}.}
\begin{eqnarray}\label{powerspectrumdefn} \langle\zeta_{\bkone}\zeta_{\bktwo}\rangle &\equiv&
\picube\,
\sdelta{\bkone+\bktwo}\frac{2\pi^2}{k_1^3}\calP_{\zeta}(k_1) \, , \\
\langle\zeta_{{\mathbf k_1}}\,\zeta_{{\mathbf k_2}}\,
\zeta_{{\mathbf k_3}}\rangle &\equiv& \picube\, \sdelta{{\mathbf
k_1}+{\mathbf k_2}+{\mathbf k_3}} B_\zeta( k_1,k_2,k_3) \,. \end{eqnarray}

From this, we can define three quantities of key observational interest, respectively the spectral
index, the tensor-to-scalar ratio and the non-linearity parameter
\bea n_{\zeta}-1&\equiv& \frac{\partial \log\calP_{\zeta}}{\partial\log k}, \\ \label{r}
r&=&\frac{\calP_T}{\calP_{\zeta}}=\frac{8\calP_*}{\Mp^2\calP_{\zeta}}, \\
\fnl&=&\frac56\frac{k_1^3k_2^3k_3^3}{k_1^3+k_2^3+k_3^3}
\frac{B_{\zeta}(k_1,k_2,k_3)}{4\pi^4\calP_{\zeta}^2}. \label{fnldefn} \eea
Here, $\calP_*$ is the
power spectrum of the scalar field fluctuations and $\calP_T=8\calP_*=8H_*^2/(4\pi^2\Mp^2)$ is the
power spectrum of the tensor fluctuations. 
Both the spectra are calculated at the end of inflation and we will ignore any evolution after this
time. As defined above, $\fnl$ is shape
dependent, but it has been shown that the shape dependent part is much less than unity 
\cite{Maldacena:2002vr,Seery:2005gb}.
The ideal CMB experiment is only expected to reach a precision of $\fnl$ around unity 
\cite{Komatsu:2001rj}, 
so we will calculate the shape independent part of $\fnl$, denoted by $\fnlf$ in
\cite{VW,Choi:2007su}.
  Whenever the non-Gaussianity is large, $|\fnl|\gtrsim1$
  as in the examples
  we are going to consider, we can
associate $\fnlf\simeq\fnl$. This ($k$ independent) part of $\fnl$ and the spectral index can be
calculated by the $\delta
N$ formalism,
\bea
{\cal P}_\zeta&=&\sum_I N_{,I}^2 {\cal P}_*,\label{spectrum} \\
n_{\zeta}-1&=& -2\epsilon^* +
\frac{2}{H}\frac{\sum_{IJ}\dot{\vp}_J N_{,JI}N_{,I}}{\sum_K N_{,K}^2},\label{index}
\\
\fnlf&=&\frac56 \frac{\sum_{IJ}N_{,IJ}N_{,I}N_{,J}}{\left(\sum_I N_{,I}^2\right)^2}.\label{deffnl}
\eea

\noindent
where $\epsilon_\star\,\equiv\,-\dot{H}_\star/H_\star^2$,
and $H_\star\,=\,H(t=t_\star)$. 
In the next sections, we will find  expressions for
the previous quantities, valid also beyond a slow-roll
approximation,
for certain classes of multi-field inflationary models.

\section{The method}\label{themethod}

In this section we present a method which allows us to obtain simple
and compact expressions for the spectral index $n_\zeta$
and the parameter $\fnl^{(4)}$, in terms of quantities associated 
with the homogeneous solution under consideration. The main
advantage of this method is that we do not 
rely on any approximation in order to find our results.
The final formulae then allow one to 
analytically study interesting regimes in field
space, that cannot be probed within 
the slow-roll approximation, usually implemented 
when studying the evolution of perturbations during
inflation.

We start in section  \ref{homeqs}
by reformulating the equations of motion as first
order Hamilton-Jacobi equations, following \cite{SB}. By implementing
a suitable Ansatz, this formulation  allows one to define two quantities that play an important
role for studying the evolution of fluctuations, that we
start discussing from section  \ref{exactev}.

\subsection{The homogeneous equations}\label{homeqs}

We consider a system  of two scalar fields $\phi$
and $\chi$ with canonical
kinetic terms, minimally coupled to gravity. 
The action, setting $M_{Pl}^2=8\pi G=1$, where $G$ is Newton's gravitational constant, is

\be
S\,=\,\int d^4 x
\sqrt{-g}\left[\frac{R}{2}-\frac12g^{\mu\nu}\partial_\mu\phi\partial_\nu\phi 
-\frac12g^{\mu\nu}\partial_{\mu}\chi\partial_\nu\chi-W(\phi,\chi) \right]. \label{action}
\ee
We focus on homogeneous, classical trajectories,
adopting a FRW Ansatz for the metric and flat three dimensional
spatial slices:
\be
d s^2\,=\,-d t^2+a(t)^2\,\left(d x_1^2+d x_2^2+d x_3^2\right)\,.
\ee
Defining the Hubble parameter as $H \equiv \frac{\dot{a}}{a}$,
the equations
of motion are

\bea
H^2 &=& \frac{1}{3}\,W(\phi,\chi)
+\frac{2}{3 }\,\left[
\left( \frac{\partial H}{\partial \phi} \right)^2+
\left( \frac{\partial H}{\partial \chi} \right)^2
 \right]\,, \label{hubeq1}\\
 \dot{\phi} &=& -2\,\frac{\partial H}{\partial \phi}
 \hskip1cm,\hskip1cm 
  \dot{\chi} \,=\, -2 \,\frac{\partial H}{\partial \chi}\,.
 \label{scaleq1} 
\eea

\smallskip

We concentrate on solutions satisfying
 the following Ansatz for the Hubble parameter $H(\phi, \chi)$:
 \be\label{anshub}
H(\phi,\,\chi)\,=\,H^{(1)}(\phi)+ H^{(2)}(\chi)\,,
\ee 
 that is, we demand that it 
can be split  as a sum~\footnote{Hence we follow an  analysis 
related to \cite{VW}. In an analogous
way, one 
could also consider a situation in which the Hubble 
parameter splits as a {\it product} of single-field pieces \cite{Choi:2007su}.}
of   two
pieces, each one depending on a single field~\footnote{
In principle, the Hubble parameter $H$ is a function 
$H(\vp^{(k)}(t),\,t)$, with $\vp^{(1)}=\phi$ and $\vp^{(2)}=\chi$.
However, by means of eq. (\ref{hubeq1}) and
(\ref{scaleq1}), we have
$$\left(
\frac{\partial H^2 }{\partial t}
\right)_{| \vp^{(k)}=const}\,=\,\frac43 
\frac{\partial H  }{\partial \vp^{(k)}}
\left[\frac{\partial^2 H  }{\partial \vp^{(k)}\,\partial t}
\right]_{| \vp^{(k)}=const}\,=\,-
\frac23 
\frac{\partial H  }{\partial \vp^{(k)}}
\left[\frac{\partial^2 \vp^{(k)}   }{\partial t^2}
\right]_{| \vp^{(k)}=const}\,=\,0
$$
So that we have always 
$H\,=\,H(\vp^{(k)}(t))$ without additional dependence
on time \cite{SB}.}.
 Notice that this  is different with  respect to the requirement
 of separable potentials usually done in the literature \cite{VW}.
 Indeed, given that the Hubble parameter appears squared
 in the left hand side of eq. (\ref{hubeq1}), it is easy to see
 that the potential cannot be written as a sum of pieces,
 each depending on  a single field.

\smallskip

This separability Ansatz allows to define two quantities 
that characterise each homogeneous trajectory
  that satisfies eqs (\ref{hubeq1})-(\ref{scaleq1}).
The first is the number of  e-foldings at time $t_c$,
that  can be expressed as
 \be
 N({t}_c, t_\star)
 \,=\,\int_{t_\star}^{{t}_c}\,H({t})\,d {t}
 \,=\,-\frac12
 \int_{\phi_\star}^{{\phi}_c}
 \,\frac{H^{(1)}}{H^{(1)}_{,\,{\phi}}}\,d {\phi}-
 \frac12
 \int_{\chi_\star}^{{\chi}_c}
 \,\frac{H^{(2)}}{H^{(2)}_{,\,{\chi}}}\,d {\chi}\,,
 \ee
where we define
$ H_{,\,\phi}\,\equiv\,
\frac{\partial H(\phi)}{\partial \phi}$ and analogously for $\chi$.
The second is a quantity ${\cal C}$, conserved along each
 homogenous trajectory, 
\be\label{difC}
{\cal C}\,=\,-\int \frac{d \phi}{H_{,\phi}^{(1)}}+
\int \frac{d \chi}{H_{,\chi}^{(2)}}
\,,\ee
whose definition resembles an analogous  quantity
used  in 
\cite{VW}.

\smallskip

While the number of e-foldings $N$ parameterises 
the evolution along a given homogeneous trajectory,
  ${\cal C}$ characterises
the motion orthogonal to it.  
We will see these quantities are
useful in order to analyse the evolution of
fluctuations at superhorizon scales, in a way
similar to what is done in 
\cite{VW}.  The main advantage of our Ansatz (\ref{anshub}) 
for the Hubble parameter is that the evolution
of perturbations can be studied without
having to rely on any approximation such as
slow roll. The formulae that we are going
to determine, then, can be applied also in regimes
where slow-roll conditions are violated: this fact
will be of crucial importance for the applications
that we discuss in the second part of the paper.

\subsection{Evolution of perturbations}\label{exactev}

As explained in section \ref{briefreview}, we intend
to follow the evolution of the curvature
perturbation between an initially flat slice at
time $t_\star$ (corresponding to Hubble exit) 
up to a final uniform density hypersurface at time $t_c$.
 To calculate the scalar spectral index 
 and the $\fnl$ parameter, as in eqs.
 (\ref{index}) and  (\ref{deffnl}), we have
 to
  compute
 the first and second derivatives of the number of 
 e-foldings along the initial values for the fields $\phi_\star$
and $\chi_\star$. In order to do this, we 
use the properties
of the functions $N(t_\star,\, t_c)$ and $C(\phi_\star,\,\chi_\star)$,
and 
 we closely follow the arguments
developed in \cite{VW}  transposing them
to our case of separable 
Hubble parameter.
We can write

\bea
d\,N&=&\frac12\,\left[ 
\frac{ H^{(1)}_\star}{ H^{(1)}_{,\,\phi_\star}}
-\frac{ H^{(1)}_c}{ H^{(1)}_{,\,\phi_c}}\,\frac{\partial
\phi_c}{\partial \phi_\star}-
\frac{ H^{(2)}_c}{ H^{(2)}_{,\,\chi_c}}\,\frac{\partial
\chi_c}{\partial \phi_\star}
\right]\,d \phi_\star\nonumber\\
&+&
\frac12\,\left[ 
\frac{ H^{(2)}_\star}{ H^{(2)}_{,\,\chi_\star}}
-\frac{ H^{(2)}_c}{ H^{(2)}_{,\,\chi_c}}\,\frac{\partial
\chi_c}{\partial \chi_\star}-
\frac{ H^{(1)}_c}{ H^{(1)}_{,\,\phi_c}}\,\frac{\partial
\phi_c}{\partial \chi_\star}
\right]\,d \chi_\star\,.
\eea
We then need the derivatives of $\phi_c$, $\chi_c$
along the initial values of the fields. The
following relations hold
\bea
d \phi_c &=& \frac{d \phi_c}{ d {\cal C}}\,\left( \frac{\partial
 {\cal C}}{\partial
\phi_\star}
\,d \phi_\star+
\frac{\partial
 {\cal C}}{\partial
\chi_\star}
\,d \chi_\star
 \right)\,, \\
 d \chi_c &=& \frac{d \chi_c}{ d {\cal C}}\,\left( \frac{\partial
 {\cal C}}{\partial
\phi_\star}
\,d \phi_\star+
\frac{\partial
 {\cal C}}{\partial
\chi_\star}
\,d \chi_\star
 \right)\,,
\eea
since $\phi_c$ and $\chi_c$
are functions of the conserved quantity ${\cal C}$.
 Also, we have

\bea
\frac{\partial {\cal C}}{\partial \phi_\star} &=&-\frac{1}{ H^{(1)}_{,\,\phi_\star}}\hskip0.6cm,\hskip0.6cm
\frac{\partial {\cal C}}{\partial \chi_\star} \,=\,\frac{1}{ H^{(2)}_{,\,\chi_\star}}\,.
\eea
By definition, the time $t=t_c$ corresponds
to a hypersurface of constant energy density, so that
$H(\phi_c,\,\chi_c)\,=\,const$ in a flat universe.
 This implies

\be\label{difhyp}
\frac{d \phi_c}{d {\cal C}}\,
H^{(1)}_{, \phi_c}\,+\,
\frac{d \chi_c}{d {\cal C}}\,
H^{(2)}_{, \chi_c}\,=\,0
\ee

On the other hand, differentiating 
 condition (\ref{difC})
along ${\cal C}$ one gets

\be\label{difcon}
1\,=\,-\frac{d \phi_c}{d {\cal C}}\,\frac{1}{H^{(1)}_{,\phi_c}}+
\frac{d \chi_c}{d {\cal C}}\,\frac{1}{H^{(2)}_{,\chi_c}}\,.
\ee
By plugging (\ref{difcon}) inside (\ref{difhyp})
one can extract $$\frac{d\phi_c}{d {\cal C}}\,=\,
-\frac{H^{(1)}_{,\phi_c}\,\left(H^{(2)}_{,\chi_c}\right)^2}{\left(H^{(1)}_{,\phi_c}
\right)^2
 +\left(H^{(2)}_{,\chi_c}\right)^2}\,.
$$ 
This, together with (\ref{difC}),
is enough  to get 
$$ \frac{\partial \phi_c}{\partial \phi_\star}
\,=\,
\frac{d \phi_c}{d {\cal C}}\,
\frac{\partial C}{\partial \phi_{\star}}\,=\,\frac{1}{H^{(1)}_{,\phi_c} H^{(1)}_{,\phi_\star}}
\,\left( \frac{1}{\left(H^{(1)}_{,\phi_c}\right)^2} + 
 \frac{1}{\left(H^{(2)}_{,\chi_c}\right)^2} 
\right)^{-1}
$$
and analogous formulae for the other derivatives of the fields. 
At this point it is convenient to define the following quantities:
\bea
\delta^{\phi}&=&\left( \frac{H^{(1)}_{,\phi}}{H}\right)^2
\hskip0.6cm,\hskip0.6cm\delta^{\chi}\,=\,\left( \frac{H^{(2)}_{,\chi}}{H}\right)^2\,,
 \\
\gamma^{\phi}&=& \frac{H^{(1)}_{,\phi \phi}}{H}
\hskip1.2cm,\hskip1.2cm\gamma^{\chi}\,=\,
\frac{H^{(2)}_{,\chi \chi}}{H}\,,
\eea
and $\delta\,=\,\delta^{\phi}+\delta^{\chi}$.  
Although their definition resembles the corresponding
 one for  the usual slow roll parameters
$\epsilon$ and $\eta$, they do {\it not} coincide with them
 when taking  a slow
roll limit~\footnote{\label{footnotesr} It is nevertheless
simple to work out the relation with the slow roll parameters,
defined as $$ \epsilon^{\phi}\,\equiv\frac12\,\left( \frac{W_{,\phi}}{W}\right)^2\hskip0.7cm\eta^{\phi\phi}\,=\,
\frac{W_{,\phi\phi}}{W}$$ and analogously for $\chi$. In a slow-roll
regime, one finds
$$\delta^\phi\,\simeq\,\frac{\epsilon^\phi}{2}\hskip0.7cm
\gamma^{\phi}\,\simeq\,\frac{\eta^{\phi\phi}-\epsilon^\phi}{2} \,.$$}. 
We
assume that the quantities $\delta$
 and $\gamma$ are much smaller than unity at Hubble exit
 $t=t_\star$, in order
to have a reliable expansion for $\delta N$ 
as explained in section  (\ref{briefreview}). During inflation, by definition,
we have to ensure that the quantity
  $\epsilon_H\,\equiv\,-\dot{H}/H^2$ is smaller than
  one. On the other hand,
\be
\epsilon_H\,=\,-\frac{1}{H^2}\,\left( H_{,\phi}^{(1)}\,\dot{\phi}+
H_{,\chi}^{(2)}\,\dot{\chi}\right)\,=\,2\,\delta\,,
\ee
implying  that $\delta$ must remain small during the inflationary
period. The quantities $\gamma$ can however  become
large during inflation: this property will 
be very important in the following discussion. 

\smallskip

By means of these quantities, we can write
\bea
 \frac{\partial \phi_c}{\partial \phi_\star}
&=&
\frac{H_c}{H_\star}
\,\frac{\delta_c^\chi}{\delta_c}\,\left(\frac{\delta_c^\phi}{\delta_\star^\phi}\right)^{\frac12}
%
\hskip
1cm,\hskip1cm 
 \frac{\partial \phi_c}{\partial \chi_\star}
\,=\,-
\frac{H_c}{H_\star}
\,\frac{\delta_c^\chi}{\delta_c}\,\left(\frac{\delta_c^\phi}{\delta_\star^\chi}\right)^{\frac12}\,,
 \\
  \frac{\partial \chi_c}{\partial \phi_\star}
&=&-
\frac{H_c}{H_\star}
\,\frac{\delta_c^\phi}{\delta_c}\,\left(\frac{\delta_c^\chi}{\delta_\star^\phi}\right)^{\frac12}
%
\hskip
0.8cm,\hskip1cm 
 \frac{\partial \chi_c}{\partial \chi_\star}
\,=\,
\frac{H_c}{H_\star}
\,\frac{\delta_c^\phi}{\delta_c}\,\left(\frac{\delta_c^\chi}{\delta_\star^\chi}\right)^{\frac12}\,.
\eea
We now have all the instruments to compute the derivatives of the
number of e-foldings. The first derivatives result
\bea\label{Np}
\frac{\partial N}{\partial \phi_\star}=\frac{1}{2 \sqrt{
\delta_\star^\phi}}\,\frac{H_{\star}^{(1)}+Z_{c}}{H_{\star}}\,,\qquad
\frac{\partial N}{\partial \chi_\star}=\frac{1}{2 \sqrt{
\delta_\star^\chi}}\,\frac{H_{\star}^{(2)}-Z_{c}}{H_{\star}}\,,
\eea
where 
\be
Z_c=\frac{H_c^{(2)}\,\delta_c^{\phi}-
H_c^{(1)}\,\delta_c^{\chi}
}{\delta_c}\,.
\ee 
This quantity $Z_c$  controls the evolution of the curvature perturbation at superhorizon
scales, as expected in multi-field inflationary models. 

The calculation of the
 second derivatives yields 
 \bea\label{Npp}
 \frac{\partial^2 N}{\partial \phi_\star^2}&=&
 \frac12-\frac{\gamma_\star^\phi}{2\,\delta_\star^\phi}\,
 \frac{H_\star^{(1)}+Z_c}{H_\star}+\frac{1}{2 H_\star\,\sqrt{\delta_\star^\phi}}\,\frac{\partial
Z_c}{\partial \phi_\star}
 \,,\\ 
  \frac{\partial^2 N}{\partial \chi_\star^2} &=&
   \frac12-\frac{\gamma_\star^\chi}{2\,\delta_\star^\chi}\,
 \frac{H_\star^{(2)}-Z_c}{H_\star}-\frac{1}{2 H_\star\,\sqrt{\delta_\star^\chi}}\,\frac{\partial
Z_c}{\partial \chi_\star}
 \,,\\
   \frac{\partial^2 N}{\partial \chi_\star\,\partial\phi_\star} &=&
   \frac{1}{2 H_\star\,\sqrt{\delta_\star^\phi}}\,\frac{\partial Z_c}{\partial \chi_\star}\,=\,-
    \frac{1}{2 H_\star\,\sqrt{\delta_\star^\chi}}\,\frac{\partial Z_c}{\partial \phi_\star}\,.
\label{Npc} \eea
We now have all the necessary quantities for evaluating
the expressions discussed in section \ref{briefreview}.

\subsection{Curvature perturbation
and  non-Gaussianities during inflation}

We can use eqs. (\ref{index}) and (\ref{deffnl}) to calculate
the scalar spectral index during inflation, as well as the
non-linear parameter $\fnl$. In order to do this, 
it is convenient to introduce two new variables
\be
u\equiv \frac{H_\star^{(1)}+Z_c}{H_\star} \hskip1cm,\hskip1cm
v\equiv \frac{H_\star^{(2)}-Z_c}{H_\star}\,,
\ee
with the property  that $u+v=1$. 
By means of these and of the results
of the previous section, the spectral index
for the curvature can
 be straightforwardly calculated from eq. (\ref{index}), giving
\be\label{nzeta}
n_{\zeta}-1\,=\,-  4\delta_\star-\,4\,\frac{u\left( 
1-\frac{\gamma_\star^\phi}{ \delta_\star^\phi}
u \right)+v \left(
1-\frac{\gamma_\star^\chi}{ \delta_\star^\chi}
v
\right)}{\frac{u^2}{\delta_\star^\phi} + 
\frac{v^2}{\delta_\star^\chi} 
}\,.
\ee
Notice that, 
since it depends on the parameters $u$ and $v$, it
changes with time also beyond horizon exit, due
to the correlation  between  entropy and adiabatic
perturbations. On the other hand, if $\delta_\star$
and $\gamma_\star$ are small, the spectral index 
normally remains close to one during the entire inflationary
period.

\smallskip

We pass then to calculate $\fnl^{(4)}$,
 the parameter on which we are more interested in. 
 We start by introducing the quantity
\be
{\cal A}\,=\,-\frac{H_c^2}{H_\star^2}\,\frac{\delta_c^\phi \delta_c^\chi}{
\delta_c}\,\left( \frac12-\frac{\gamma^{ss}_c}{\delta_c}\right)\,,
\label{defcalA}\ee
where we define
\be
\gamma^{ss}\,=\,\left(\delta^\chi \gamma^\phi+ 
\delta^\phi \gamma^\chi\right)/
\delta\,.
\ee 
One can then express the value for the
 $f_{NL}^{(4)}$ by plugging  the previously
 obtained results in formula (\ref{deffnl}),
 finding~\footnote{Notice the formal similarity
 with equation (68) in \cite{VW}.  The meaning of the 
 various quantities, however, is different with respect
 to that paper. Various numerical factors are also different.}

 \be\label{keyresult}
\frac{6}{5}\,f_{NL}^{(4)}\,=\, 2\frac{
\frac{u^2}{\delta_\star^\phi}\,\left( 1
-\frac{\gamma^{\phi}_\star}{ \delta_\star^\phi}
\,u \right)+
\frac{v^2}{\delta_\star^\chi}\,\left( 1
-\frac{\gamma^{\chi}_\star}{ \delta_\star^\chi}
 \,v\right)
+ 2\left( \frac{u}{\delta_\star^\phi}
-  \frac{v}{\delta_\star^\chi}
\right)^2\,{\cal A}
}{\left( \frac{u^2}{\delta_\star^\phi}
+  \frac{v^2}{\delta_\star^\chi}
\right)^2}\,.
\ee
This is one of the main results of our paper. We reiterate that this formula is exact and is not
based on a slow-roll expansion.  
Notice that  the first two terms  of eq. (\ref{keyresult}) 
 are suppressed by $\delta_\star^\phi$ or $\delta_\star^\chi$,  
  and these quantities are less than unity since they are calculated
  at horizon exit. Only the last term proportional to ${\cal A}$
  is not suppressed by quantities evaluated at horizon crossing.
  It depends on the quantity $\gamma_c^{ss}$, that can become
  much bigger than one during inflation, rendering the
  size of ${\cal A}$ large enough to enhance
   $f_{NL}$ to  observable values.
 Our formalism allows to study 
this regime in a reliable, analytical fashion, contrary to an analysis relying
on a slow-roll approximation. It is also possible to have a large $f_{NL}$ while all slow-roll
parameters are small, if the ratio of some of the parameters is very large and the background
trajectory curves in an appropriate way \cite{Byrnes:2008wi} (see also
\cite{Alabidi:2006hg,Byrnes:2008zy}). We note in agreement with \cite{VW} that if one of the fields
has reached a minimum so $\dot{\phi}=0$ or $\dot{\chi}=0$ at the end of inflation then
$\mathcal{A}=0$ and the non-Gaussianity at the end of inflation will be small. Therefore any model
of inflation with a separable potential or separable Hubble factor with a large non-Gaussianity
present at the
end of inflation must have both fields still evolving, and therefore the presence of isocurvature
modes. It would therefore also be interesting to study the evolution of the perturbations after
inflation, to see if this leads to an important change in the observables \cite{Choi:2008et}.

\smallskip

In the next section, we apply our general results to various
examples of inflationary trajectories, showing in concrete
examples 
that our formulae can analytically probe regions
 in field space where the non-linear parameter 
 $\fnl^{(4)}$ is enhanced to large values.

\section{Exact solutions}\label{exactsolutions}

Our system of first order equations, discussed
in section \ref{homeqs}, also allows one to find new exact inflationary
trajectories in the multi-field case, provided that the scalar
potential satisfies some conditions. There are very few examples of multiple field exact solutions
 in the literature \cite{Sasaki:2007ay}. In this section, we
present two classes of new exact solutions, with the aim to 
study the evolution of non-Gaussianities in each example. In the 
first example, section \ref{quadpot}, we analyse a quadratic potential,
showing that in this case the non-gaussian
parameter $\fnl^{(4)}$ 
  {\it always} remains small
  during the inflationary regime.
   In the second example, 
section \ref{exppot}, we consider a combination
of exponential potentials. In this case, non-Gaussianities
can get enhanced to observable values, 
 depending on the choice of the parameters
 in the potential.

\subsection{Quadratic potential: no enhancement of $\fnl$}
\label{quadpot}

We first present an example of exact configuration
that does {\it not} lead to enhanced 
non-Gaussianities, since 
the quantity  ${\cal A}$ defined
in eq. (\ref{defcalA}) does not increase much during inflation.

Consider a quadratic potential
\be
W\,=\, A  +B \,\phi +C\,\chi+D\,\phi^2+E\,\chi^2+F\,\phi \chi
\ee
where the constants $B$, $C$ and $F$ are related to the other
quantities by

\bea
B^2&=&\frac{8\,D}{3}\,\left( \frac32 A +D+ E \right) \,,\\
C^2&=&\frac{8\,E}{3}\,\left( \frac32 A +E+D  \right)\,,\\
F^2&=&4\,D E\,.
\eea
The equations of motion
can be solved choosing a Hubble parameter of the form
\be
H\,=\,H_0+H_1 \phi+ H_2 \chi\,,
\ee
with 
\bea
H_0^2=\frac29\,\left( \frac32 A +E +D  \right)\,, \qquad H_1^2=\frac{D}{3}\,, \qquad
H_2^2=\frac{E}{3}\,.
\eea

Setting for simplicity, and without loosing
generality,   $t_\star=0$,
the equations have the solution 
\bea\label{solffields}
\phi(t)&=& \phi_\star-2 H_1 t \,, \qquad \chi(t)= \chi_\star-2 H_2 t \,,\\
a(t) &=& a_0 \,e^{\left( H_0 +H_1\,\phi_\star+H_2 
\,\chi_\star\right) t-(H_1^2+H_2^2) t^2}\,.
\eea
Notice that, although the solution is well defined for all $t$, at 
sufficiently large times the Hubble parameter becomes negative and
the universe starts contracting. This point is, in any case, well 
beyond the end of the inflationary era, and we do not need 
to discuss it in our analysis.
Inflation ends when the parameter $\epsilon_H$ 
given by 
\be\epsilon_H(t)\,=\,-\dot{H}/H^2\,\,=\,\frac{2\left( 
H_1^2+H_2^2\right)}{\left[
H_0 +H_1\,\phi_\star+H_2 
\,\chi_\star-2 (H_1^2+H_2^2) t
 \right]^2}
=\,2\, \delta \label{nfpot}\ee
becomes unity. Notice that the equations of motion, and eq. (\ref{nfpot}),
allows one to rewrite the expression for the potential
in a suggestive form
\bea
W= 3 H^2-2\left( H_1^2+H_2^2\right) =3 H^2\,\left( 1-\frac{\epsilon_H}{3}\right)
\eea
ensuring that during inflation this quantity remains positive definite.

\smallskip

 In order to rely on the expansion for the number of e-foldings
 discussed in section \ref{briefreview},  
 we  impose that 
the quantities $\delta_\star^{\phi,\,\chi}$ 
are small. From
eq. (\ref{nfpot}), we learn that
this is ensured as long as $H_1,\,H_2\,\ll H_0$. 
 Notice
that the $\gamma$
parameters in this example vanish  identically: 
this simplifies the expression for the spectral index,
that reduces to
\be
n_{\zeta}\,=\,1-4\delta_\star-4\,\frac{\delta_\star^\phi \delta_\star^\chi 
}{
u^2 \delta_\star^\chi+v^2 \delta_\star^\phi  
}
\ee
so $n_\zeta-1$ is proportional  to terms suppressed
by small $\delta_\star$'s parameters. 
Another consequence of having
$\gamma^{\phi,\,\chi}\,=\,0$, is the fact
that the quantity ${\cal A}$ cannot
increase much during inflation.
So, following the discussion
of the previous sections, we do not expect to find any enhancement 
of non-Gaussianities. Indeed, it
is straightforward to get  the following expressions
for  the parameters $u$ and $v$: $u = \delta_\star^\phi/\delta_\star$,
 $v = 
\delta_\star^\chi/\delta_\star$.
A simple calculation then  provides
\bea
\frac65 \, f_{NL}&=&\frac{2\,\delta_\star^\phi \delta_\star^\chi}{
\left(\delta_\star^\chi \,u^2+\delta_\star^\phi \,v^2\right)
}-2\, \frac{\left(\delta_\star^\chi \,u-\delta_\star^\phi 
\,v\right)^2}{\left(\delta_\star^\chi \,u^2+\delta_\star^\phi 
\,v^2\right)^2}\,\frac{\delta_\star^\phi \delta_\star^\chi}{\delta_\star}\,
\nonumber\\
&=& 2\, \delta_\star,
\eea
that  then remains small and constant during inflation.
This is not surprising since, from eq. (\ref{solffields}),
 one learns
that the inflationary trajectory follows a straight line 
in field space. 
 To finish the discussion of this example, notice that spectral
 index and $f_{NL}$ are connected by the relation
 $n_\zeta-1\,=\,-\frac{24}{5} f_{NL}$.

\subsection{Exponential potential: enhancement
of $\fnl$ towards the end of inflation} \label{exppot}

We discuss here an exact inflationary trajectory,
for a potential  based on a combination
of exponentials that is instead able to produce
large values of $\fnl$ at the end of
inflation. Noticeably,  a potential with a similar 
form 
may be motivated in string theory, in the context
of K\"ahler moduli inflation \cite{kahlerinfl}.
We start by presenting the inflationary trajectory,
and showing how it can enhance the non-linearity
parameter $\fnl$. We discuss the physical
interpretation of our findings at the end of this
section.

The potential we consider is
\be
W(\phi,\chi)\,=\,U_0\,\left( 1-A_1 e^{-\alpha \phi}
+A_2  e^{-2 \alpha  \phi}
-B_1 e^{-\beta \chi}+
B_2 e^{-2\beta \chi}+
\frac{ A_1  B_1}{2}\,e^{-\alpha \phi-\beta \chi}\right)\,,
\ee
where the parameters
$\alpha$ and $\beta$
satisfy 
\bea\label{alphabetaconstraints}
\alpha^2 =\frac32-\frac{6 A_2}{A_1^2}\,, \qquad \beta^2 =\frac32-\frac{6 B_2}{B_1^2}\,.
\eea
Since one may freely redefine the initial field values, we will assume that $\alpha, \beta>0$ without loss of generality \cite{Battefeld:2009ym}, and we also assume that $A_1,A_2>0$ \footnote{We did not make these positivity assumptions in versions 1 and 2 of 
this paper, see the ``Note added'' at the end of the
conclusions.}.

The equations of motion discussed
in section \ref{homeqs} admit the following 
 solution  for $H$:
\be
H\,=\,\sqrt{\frac{U_0}{3}}\,\left(1-\frac{A_1}{2}
 e^{-\alpha \phi}-\frac{B_1}{2} e^{-\beta \chi}\right)\,.
\ee
The overall factor to the potential $U_0$ can be freely chosen so that the amplitude of the scalar
power spectrum matches the observed amplitude of perturbations in the CMB.
 
The solutions for the scalar fields read 
(from now on $H_0=\sqrt{U_0/3}$ and we have again set $t_{\star}=0$)

\bea
\phi=\frac{1}{\alpha}\,\ln{\left[e^{\alpha \phi_{\star}}
-A_1 \,\alpha^2\,H_0\,t
\right]}\,, \qquad \chi=\frac{1}{\beta}\,\ln{\left[e^{\beta \chi_{\star}}
-B_1 \,\beta^2\,H_0\,t
\right]}\,.
\eea

The scale factor results:
\be
a(t)\,=\,a_0\,\left(e^{\alpha\phi_{\star}}-
  \alpha^2\, A_1\,H_0\,t
\right)^{\frac{1}{2 \alpha^2}}\,\left(e^{\beta\chi_{\star}}-
  \beta^2\, B_1\,H_0\,t
\right)^{\frac{1}{2 \beta^2}}\,e^{H_0\,t}\,.
\ee
Notice that the solution becomes singular at late times, when the scale 
factor vanishes and the field values diverge. The singularity, on the other hand, occurs well after 
inflation ends, and we will not need to discuss it in our analysis.
The quantities $\delta$ and $\gamma$
read in this case
\bea
\delta^\phi&=&\frac{H_0^2}{ H^2}\,\frac{\alpha^2 A_1^2}{4}\,
e^{-2 \alpha \phi}\hskip0.5cm,\hskip0.5cm
\delta^\chi\,=\, \frac{H^2_0}{ H^2}\,\frac{\beta^2 B_1^2}{4}\,
e^{-2 \beta \chi}\,,\\
\gamma^\phi&=&
-\frac{H_0}{ H}\,\frac{\alpha^2 A_1}{2}\,
e^{- \alpha \phi}\hskip0.5cm,\hskip0.5cm
\gamma^\chi\,=\, -\frac{H_0}{ H}\,\frac{\beta^2 B_1}{2}\,
e^{- \beta \chi}\,.
\eea
Useful relations are
\bea
\gamma^{\phi}&=&-\alpha\,\sqrt{\delta^\phi}\hskip0.5cm,\hskip0.5cm
\gamma^{\chi}\,=\,-\beta\,\sqrt{\delta^\chi} \,,\label{relgam}\\
H&=&\frac{H_0}{1-\frac{\gamma^\phi}{\alpha^2}-
\frac{\gamma^\chi}{\beta^2}} \label{relH}\,.
\eea
Notice that inflation occurs while
\be
\epsilon_H\,=\,2 \delta^\phi+2 \delta^\chi\,=\,2 \left(\frac{\gamma^\phi}{
\alpha}\right)^2+2 \left(\frac{\gamma^\chi}{
\beta}\right)^2\,\le\,1\,.
\ee
In this example $\gamma^{\phi}$ and $\gamma^{\chi}$ can become much larger than unity at the end of
inflation if $\alpha$ and/or $\beta$ are much greater than one. This corresponds to a break down in
slow roll, although $\epsilon_H$ remains smaller than unity during the inflationary era.

We can assume that at the initial time $t=t_\star$,
 the gamma parameters
are smaller than $\alpha$ and $\beta$,
so  that $H_\star\simeq H_0$. Using the formalism
developed in section \ref{themethod},  we call 
$$ H^{(1)}(\phi)\,=\,\frac{H_0}{2}-\frac{A_1\,H_0}{2}\,e^{-\alpha \phi}
\,=\,
\frac{H_0}{2}\,\left(1-
 \frac{2\,\sqrt{\delta_c^\phi}}{\alpha}\,\frac{H_c}{H_0}\right)
\hskip0.5cm,\hskip0.5cm H^{(2)}(\chi)\,=\,H(\phi,\chi)- H^{(1)}(\phi)
$$
The number of e-folding at time $t_c$ reads

\bea\label{efoldpar}
N(\phi_c,\chi_c)&=&\frac{1}{2 \alpha}\,\left(\phi-\phi_{\star}\right)
+\frac{1}{2 \beta}\,\left(\chi-\chi_{\star}\right)
-\frac{1}{2 A_1 \alpha^2}\,\left(e^{\alpha \phi}
-e^{\alpha \phi_{\star}}\right)
-\frac{1}{2 B_1 \beta^2}\,\left(e^{\beta \chi}
-e^{\beta \chi_{\star}}\right)\nonumber \\
&=&\ln{\left(
\frac{\delta_\star^{\phi}}{\delta_c^{\phi}}
\,\frac{H^2_\star}{H^2_c}
\right)^{\frac{1}{4 \alpha^2}}}
+\ln{\left(
\frac{\delta_\star^{\chi}}{\delta_c^{\chi}
}\,\frac{H^2_\star}{H^2_c}\right)^{\frac{1}{4 \beta^2}}
}
+\frac14 \left( \frac{1}{|\gamma_\star^\phi|}+
 \frac{1}{|\gamma_\star^\chi|}
\right)-\frac{H_0}{4 H_c} \left( \frac{1}{\alpha\,\sqrt{\delta_c^\phi}}+
 \frac{1}{\beta\,\sqrt{\delta_c^\chi}}
\right) \label{efolds}\,.
\eea
Focussing on the regime of large $\alpha$
and $\beta$, in which the 
first two terms in the previous expression
are suppressed, 
the main contribution to the number of e-foldings
comes from the terms that go like $1/\gamma_\star$,
as long as $\delta_c^\phi$ or $\delta_c^\chi$ are not
too small (more on this later).

\smallskip

We now proceed  to calculate the quantities necessary
for estimating the non-linear parameter $\fnl$. One finds
that

\be
u\,=\,\frac12+\frac{\delta^\phi-\delta^\chi}{2\delta}
-\frac{\sqrt{\delta^\phi}}{\alpha}
+\frac{\sqrt{\delta^\phi \delta^\chi}}{\delta}\,
\left(
\frac{\sqrt{\delta^\chi}}{\alpha}
-
\frac{\sqrt{\delta^\phi}}{\beta}
\right)\,,
\ee
and 
\be
\gamma^{ss}\,=\,-\frac{\sqrt{\delta^\phi \delta^\chi}}{\delta}\,\left[\alpha\,\sqrt{\delta^\chi}+\beta\sqrt{\delta^\phi}
\right]\,.
\ee
The expression for ${\cal A}$, evaluated at $t=t_c$ is given by
\be
{\cal A}\,=\,-\frac{1}{\left(1-\frac{\sqrt{\delta_c^\phi}}{\alpha}-
\frac{\sqrt{\delta_c^\chi}}{\beta}\right)^2}
\,\left( \frac{ \delta_c^\phi \delta_c^\chi}{\delta_c^2}\right)^\frac32\,\left(
\alpha \sqrt{ \delta^\chi_c}+ 
\beta \sqrt{ \delta^\phi_c}+\frac{\delta_c^2}{\sqrt{ \delta_c^\phi \delta_c^\chi}}
\right)\,.
\ee
Although the quantities $\delta_c^{\phi}$ and
$\delta_c^{\chi}$ are smaller than one during inflation,
on the other hand the parameters $\alpha$ and $\beta$
can be large; this allows the quantity
 ${\cal A}$ to become large during inflation, enhancing
 the value of $\fnl^{(4)}$. We now show this with an explicit example.

\subsubsection*{A concrete example}

As a simple, concrete example, we choose a regime in which
the parameters $\alpha$ and $\beta$ are both large,
let us say larger than some arbitrary quantity $R\gg1$
with no physical meaning by itself.
 We note from (\ref{alphabetaconstraints}) that we
are therefore required to have $A_2<0$ and $B_2<0$. In this case the potential does not have a
minimum, but inflation still ends through $\epsilon_H$ growing larger than unity, and we can trust
our results in this regime. The potential would need modifications which apply after inflation in
order for reheating to take place and these modifications may provide a minimum for the potential.
However a study of this issue
is beyond the scope of
this work. We parameterise
the values of the quantities $\delta_c^\phi$ and $\delta_c^\chi$
at the end of inflation 
as 

\be\delta_c^\phi=\frac{1}{2 m}
\hskip0.4cm, \hskip0.4cm
\delta_c^\chi=\frac{1}{2 n}
\hskip0.8cm {\rm such \hskip 0.2cm that}
\hskip0.8cm
\frac{1}{m}+\frac{1}{n}=1\,.\label{fincondelta}\ee
 Suppose that the quantities $m$ and $n$
are not too big. Namely, they satisfy
the inequality $m,\,n\,\ll R^{2}$, such that
 the expression for the number of e-foldings, eq.
  (\ref{efolds}),
 is dominated by the terms proportional to the inverse
 of $\gamma_\star$. The same inequality also implies
 that $H_0\simeq H_\star \simeq H_c$. 
We choose
 $|\gamma_\star^\phi|=\frac{1}{40 p}$,  
  $|\gamma_\star^\chi|=\frac{1}{40 q}$, such that we can write
 \be
 N_{tot}=60\,\simeq\,10\, \left(p+q \right)\,.
 \ee
In order to have $60$ e-foldings, we have to choose $p+q=6$.
This fixes the initial values for the fields
 $\phi_\star$ and $\chi_\star$,
 to the values
 \bea
 e^{\alpha \phi_\star}&=& 20\, p\,\alpha^2 A_1 
 \label{inval1}\,,\\
 e^{\beta \chi_\star}&=& 20\, q\,\beta^2 B_1\,.
 \label{inval2}
 \eea

The conditions (\ref{fincondelta}) imply
\bea H_0\,t_c &=& 20 p-\sqrt{\frac{m}{2\alpha^2}}\,,\\
H_0\,t_c &=& 20 q-\sqrt{\frac{n}{2\beta^2}}\,.
\eea
Since $m$ and $n$ are much smaller than $R^2$, we
expect $p\simeq q$.
We hence have the
simple relations
\bea \delta^{\phi}_{\star}\simeq\frac{1}{4N_{tot}^2\alpha^2}, \qquad
\delta^{\chi}_{\star}\simeq\frac{1}{4N_{tot}^2\beta^2}, \qquad
\gamma^{\phi}_{\star}\simeq\gamma^{\chi}_{\star}\simeq-\frac{1}{2N_{tot}}\,. \eea
From (\ref{r}), (\ref{spectrum}) and (\ref{Np}) for the tensor-to-scalar ratio, and from
(\ref{nzeta}) for the spectral index,  we find
\bea r&\simeq&\frac{8}{N_{tot}^2}\frac{1}{\frac{\alpha^2}{m^2}+\frac{\beta^2}{n^2}}\ll10^{-3}\,, \\
n_{\zeta}-1&\simeq&-\frac{2}{N_{tot}}\simeq-0.04\,. \eea
So we have a completely negligible level of gravitational waves and a red spectral index in
agreement with present day observations \cite{wmap5}.

\begin{figure}
\begin{center}
\scalebox{1.2}{\includegraphics*{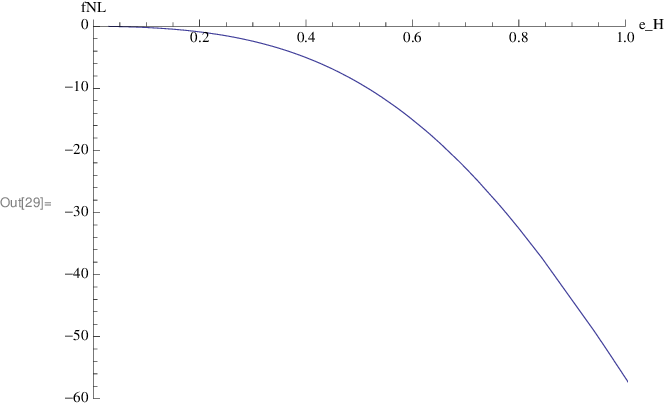}} 
\caption{Plot showing $f_{NL}$ as given by (\ref{fNLevolve}) as a  
 function of $\epsilon_H$ towards
 the end of
 inflation, for the values of the parameters given in the text. Inflation ends
 when $\epsilon_H=1$; for this example $f_{NL}\simeq-59$ at that time.}\label{plot}
 \end{center}
\end{figure}

It is interesting to notice
that for this choice of initial conditions we can
write
\begin{eqnarray}
 \phi&\simeq& 
 \frac{1}{\alpha}\,\log{\left[N_{tot}-N(t)
+\sqrt{\frac{m}{2\,\alpha^2}} \right]}+
\frac{1}{\alpha}\,\log{\left[\alpha^2 A_1\right]}\,,
\label{eqsimphi}
\\
\chi&\simeq&
\frac{1}{\beta}\,\log{\left[N_{tot}-N(t)
+\sqrt{\frac{n}{2\,\beta^2}} \right]}+
\frac{1}{\beta}\,\log{\left[\beta^2 B_1\right]}\,,
\label{eqsimchi}
\end{eqnarray}
where we used eq. (\ref{efoldpar}). $N(t)$ indicates
the number of e-folds at time $t$, and 
$N_{tot}$ corresponds to the total number of 
e-folds (that as we mentioned we take of order sixty). 
We know that the terms inside the square roots, in the
argument
of the logarithms appearing in the previous equations, are
much smaller than one since $m,\,n\ll R^2$.
This implies that 
in a regime in which $N_{tot}-N(t)\gsim1$ we can write
\be\label{straight}
\beta\,\chi(t)\,=\,\alpha\,\phi(t)-\log{\frac{\alpha^2 A_1}{\beta^2 B_1}}\,.
\ee
Since $\chi$ is proportional to $\phi$, we have a straight
trajectory in field space, implying that we cannot have 
 conversion of entropy into adiabatic  modes \cite{gordon},
 and consequently
 the curvature perturbation remains almost constant at
 superhorizon scales.  
 In this regime, for the arguments we have seen
 in the previous sections, we cannot produce
 large non-Gaussianities
 in this context.
    However, towards the end of inflation,
 $N_{tot}-N(t) < 1$  and the trajectory acquires a sharp turn, 
since 
the terms inside the square roots in eqs. (\ref{eqsimphi})
and (\ref{eqsimchi}) start to give sizeable  contributions. 
Precisely in this regime 
the curvature perturbation  evolves at superhorizon
scales, and parametrically large non-Gaussianities
can be produced, as we are going to see now.

\smallskip

We can calculate $u$ and $v$ in this case, finding
(from now on we neglect corrections of order  $1/R$)
\bea\label{uv}
u\simeq\frac12 +\frac12 \frac{n-m}{n+m}\,=\,
2 \delta_c^\phi\,, \hskip1cm 
v\simeq\frac12 -\frac12 \frac{n-m}{n+m}
\,=\,
2 \delta_c^\chi\,. 
\eea

\noindent
It is  simple to check that the  curvature
spectral index $n_\zeta$ given in eq. (\ref{index})
remains close to one during the entire inflationary period.
  
The parameter ${\cal A}$ becomes
$$ {\cal A}\,\simeq\,-\frac{1}{\sqrt{2} (m n)^\frac32}
\,\left( \frac{\alpha}{\sqrt{n}}+
\frac{\beta}{\sqrt{m}}
 \right)\,.
$$

The leading contribution to the quantity  $\fnl^{(4)}$ is
\bea\label{fNLevolve}
\frac65 \fnl &\simeq&4 
\left( \frac{  u  \,\delta_\star^\chi-
v \,\delta_\star^\phi}{
u^2 \, \delta_\star^\chi+
v^2\, \delta_\star^\phi
}\right)^2
\,{\cal A}\,.
\eea

When evaluated at the end of inflation, using (\ref{uv}) we find
\bea\label{fNLconcrete}
\frac65 \fnl &\simeq&
 - 2\sqrt{2m n}\,\left[ \frac{\alpha^2 n-\beta^2 m}{\alpha^2 n^2
 +\beta^2 m^2}
 \right]^2
\,\left( \frac{\alpha}{\sqrt{n}}+
\frac{\beta}{\sqrt{m}}
 \right) \,.\label{fNLconcrete2} \eea
We stress that this formula provide only
the dominant 
contributions to $f_{NL}$. It is valid
 in the case
$|\alpha|,|\beta| > R\gg1$, and $m,n< R^2$. 

\begin{figure}
\begin{center}
\scalebox{.7}{\includegraphics*{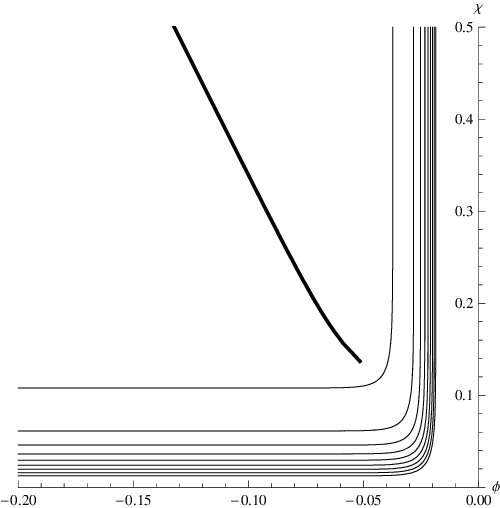}} 
$\qquad\qquad$
\scalebox{.7}{\includegraphics*{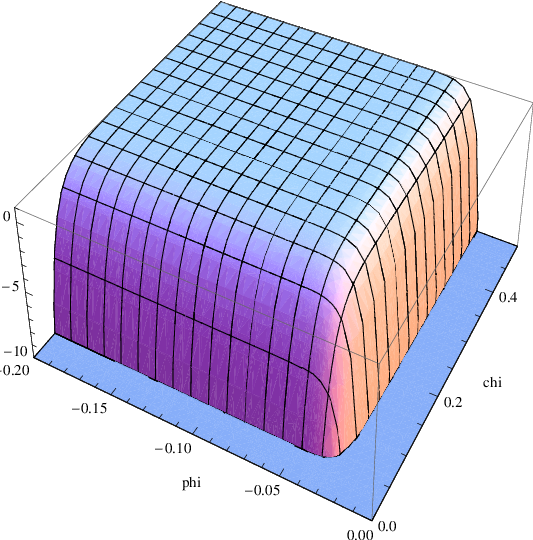}} 
\caption{Left plot shows the trajectory considered for the parameters given after
eq.~(\ref{fNLconcrete2}) superimposed on a contour plot of the potential. The square on the
trajectory indicates a point along the trajectory one $e$-folding before inflation ends as $\phi$
and $\chi$ roll towards zero. This shows that the fields roll much more quickly during the final 
stage of inflation, and the trajectory curves at the end as discussed in the text after
(\ref{straight}). The right plot shows the potential for the same parameter values. Notice that
inflation ends on the plateau long before the potential becomes negative.}\label{Traj}
 \end{center}
\end{figure}

As an explicit case,  we take $\alpha=100,\beta=20,m=6,n=6/5$ and the initial conditions to
satisfy (\ref{eqsimphi}) and (\ref{eqsimchi}) with $N_{tot}=60$. 
Then, independently of the values of
$A_1$ and $B_1$, we find $f_{NL}\simeq-58$ from the simplified formula (\ref{fNLconcrete}), and
$f_{NL}\simeq-53$ 
from the full
formula (\ref{keyresult}). See figures \ref{plot} and \ref{Traj}. Notice that the results for
$f_{NL}$ differ from
each other by quantities of order $1/R=1/\beta$ as expected by the approximations we have made.
At the end of inflation, the parameters are
$\gamma^{\phi}=29$, $\gamma^{\chi}=-13$ and this shows that the slow-roll approximation has broken
down, see footnote \ref{footnotesr}.

\smallskip

Let us end with some considerations
regarding our findings in this
concrete example. Our expression
for $\fnl$ in eq.~(\ref{fNLconcrete2})  depends
on the initial values 
%
through $p$ and $q$, and
this dependence drops out at leading order in $\alpha$.
Hence $f_{NL}$ is {\it independent} of the number of
$e$--foldings and scale independent~\footnote{For an example of a potentially scale--dependent local
$f_{NL}$ see \cite{Byrnes:2008zy}, and for a scale dependent equilateral $f_{NL}$ see for example
\cite{Chen:2005fe}. Observational prospects were recently considered in \cite{Sefusatti:2009xu}.}.
This is easy
to understand, since  we have seen that, in our
example,  non-Gaussianity
can be produced only towards the very end of inflation,
when the trajectory has a sharp turn in field space. 

Notice that our requirements of final values for the quantities
$\delta_c^\phi$ and $\delta_c^\chi$ (both much bigger than
$1/R$) imposes fine-tuning constraints on the initial 
conditions, since at leading order in $1/R$ the values
for $\gamma_\star^\phi$ and
$\gamma_\star^\chi$ must coincide
 (see eqs. (\ref{inval1})-(\ref{inval2})
and recall that $p\simeq q$). Importantly though this kind of fine-tuning does not force us to
choose a very special field value which would be inconsistent with the $\delta N$ expansion,
specifically we have checked that we satisfy $\dot{\phi}\gg H^2$ and similarly for the $\chi$ field
\cite{Creminelli:2008es}. For more discussion of this point see \cite{Byrnes:2008zy}.
\smallskip

We stress that the enhancement of non-Gaussianity occurs towards the end of inflation. What happens
just {\it after} inflation is a model dependent issue; we cannot address this question within the
approximations used in this concrete model. For our choice of parameters we notice that $f_{NL}$ is
still increasing at the end of inflation, see figure \ref{plot}, but for other choices it may start
to decrease before inflation ends. We stress that in the regime where $\alpha,\beta\gg1$ the
slow-roll parameters will necessarily become much greater than unity by the end of inflation, which
may correspond to $|\gamma^{ss}|\gg1$: this is one reason by which $f_{NL}$ can become large. Our
formalism is at least in principle suitable to study the evolution of non-Gaussianity after
inflation, when the parameter $\epsilon_H$ becomes larger than unity. We intend to address this
issue in a future publication.

\subsection*{Trispectrum}

In this regime where $\alpha\gg\beta>R$ and $m,n>1/R^2$ it is also
possible to give compact expressions for the trispectrum (4-point function) non-linearity
parameters. Unlike for the bispectrum there are two non-linearity parameters which in principle are
observationally distinguishable, since they multiply terms with different shape dependence. Using
the delta-$N$ formalism they are given by \cite{Seery:2006js,Byrnes:2006vq,Alabidi:2005qi}
\bea \tau_{NL}= \frac{\sum_{IJK} N_{,IJ}N_{,IK}N_{,J}N_{,K}}{\left(\sum_I N_{,I}^2\right)^3},
\qquad g_{NL}=\frac{25}{54}\frac{\sum_{IJK}N_{,IJK}N_{,I}N_{,J}N_{,K}}{\left(\sum_I
N_{,I}^2\right)^3}. \eea
We already have all the ingredients to give the full expression for $\tau_{NL}$, eqs.~(\ref{Np})-
(\ref{Npc}), and in principle it
is a straightforward but lengthy calculation to differentiate these to get the third derivatives as
required by $g_{NL}$. The full results are extremely cumbersome,
but  it is not hard to
 estimate the size
of these terms, in the regime we are currently studying. Indeed
 it is simple to see, from eqs.~(\ref{Np})-
(\ref{Npc}), that at leading order in $\alpha$ the term  $\partial N/(\partial\phi_{\star})$
dominates at linear order in the delta-$N$ expansion, 
while
  at second order is the term $\partial^2
N/(\partial\phi_{\star}^2)$ that dominates. With some more 
 calculations, using the methods of
Sec.~\ref{exactev},  we find that the largest term at third order, at leading order in $\alpha$ and
evaluated at the end of inflation,  is given by
\bea \frac{\partial^3 N}{\partial\phi_{\star}^3}\simeq
6\frac{\left(\sqrt{2}-2\beta\sqrt{n}\right)\sqrt{n}\left(n-m\right)}{\beta m^3 n^5}\alpha^5
N_{tot}^3. \eea
Now also using the fact that $\beta\gg1$ we can write the two terms for the trispectrum in terms of
$f_{NL}^2$ as
\bea \tau_{NL}\simeq\left(\frac65 f_{NL}\right)^2, \qquad \frac{54}{25}g_{NL}\simeq-\frac32
\frac{n-m}{m}\left(\frac65 f_{NL}\right)^2. \eea
The result for $\tau_{NL}$ is as expected for a model where a single field direction generates the
primordial curvature perturbation \cite{Byrnes:2006vq} as it is a lower bound since in general
$\tau_{NL}\geq(5 f_{NL}/6)^2$ \cite{Suyama:2007bg}. For a discussion of observational prospects for
the trispectrum see \cite{Kogo:2006kh}. We note that if $m\simeq1$ then from
(\ref{fincondelta}) $n\gg1$, so the trispectrum through $g_{NL}$ will give the dominant signal of
non-Gaussianity through a large, negative $g_{NL}$.

\bigskip

\section{Conclusions and Outlook}\label{conclusions}

We have considered models of inflation driven 
by multiple scalar fields with canonical
kinetic terms. We studied  the properties of  
 non-Gaussianity, parameterised 
  by the quantity  
  $f_{NL}$ characterising
the bispectrum, 
generated 
by the evolution of scalar perturbations on super Hubble scales.  
We developed a new method,
based on a first--order Hamilton--Jacobi formalism, 
that provides analytical expressions
for $f_{NL}$  valid in regimes
   {\it beyond} a slow--roll approximation.
   The main idea on which our arguments 
   were based is     to focus on  inflationary multi-field trajectories
  in  which the {\it Hubble rate}, and not the potential,
   is separable. 
  This approach allowed us to analytically
   study cases where the 
  non-Gaussian parameter
  $f_{NL}$ becomes large,  in regimes in which the slow--roll
parameters are enhanced well after Hubble exit, but before inflation ends.

By means of the same method, we also  found
new, exact multi-field  homogeneous inflationary
trajectories, for which a large amount of  non-Gaussianity can be generated before the
end of inflation. This allowed us to apply our general 
formulae  on a concrete example, showing how
the $f_{NL}$ parameter can grow parametrically
large as a function of quantities that characterise
a specific model. In the example we considered, the slow-roll approximation strongly breaks
down shortly before the end of inflation: $\fnl$
grows to large values, and it also
turns out to be scale independent. Furthermore the trispectrum is also large in this case, and
for some parameter values and initial conditions the first signal of non-Gaussianity could come
from a large $g_{NL}$. This is the first explicit example of a break down in slow roll after
Hubble exit giving rise to $|f_{NL}|\gg1$ at the end of inflation.

\smallskip
 
 Our method and results can be applied
  to other contexts. Other multi-field
   models of inflation, besides the ones we considered, 
   can have inflationary trajectories that satisfy our
   Ansatz of separable Hubble potential. All these models
   can be studied 
   by means of our methods, helping to analytically
   understand regimes in which $\fnl$ can be enhanced
   to observable values, and possibly
   finding new ways to enhance non-Gaussianities.
   In some cases, depending
   on the form of the potential and of the inflationary
   trajectory,
    the evolution of non-Gaussianity
   can be studied also {\it after} inflation ends, since
   our technique allows to go beyond a slow-roll approximation. For this reason it would be
especially interesting to find an example of a potential with a minimum and large non-Gaussianity,
so that we could address the question of how $f_{NL}$ continues to evolve after the end of
the inflationary period.
    Also, one
    can straightforwardly
 extend our method  to study multi-field inflationary models
  with non-canonical
 kinetic terms, first studied in \cite{spinflation}.
 This can allow to 
 further  explore the consequences  of
 having  non-canonical kinetic terms
 for the evolution
 of scalar fluctuations, besides
  what has been done so far in the literature \cite{DBI,RT}. 
  We hope to return to discuss these issues soon.
 
\indent {\bf Note added} This version supercedes the article published in JCAP. In the previous version and the published version, there was a sign error affecting $\fnl$ in Sec.~\ref{exppot}. This was due to our previous choice of $\alpha<0$ in that section, which led to a wrong sign of $\gamma^{\phi}$ and follow on errors in several equations. We thank Ewan Tarrant and Godfrey Leung for pointing this out to us. Our results now agree with a simultaneously updated version of \cite{Battefeld:2009ym}.

\section*{Acknowledgments}

We thank Nelson Nunes and David Mulryne for useful discussions
on related topics at the beginning of this project. Many thanks to
  Ki-Young Choi, Kazuya Koyama, Nelson Nunes, Filippo Vernizzi, David Wands and Ivonne Zavala for important commments
on a draft of this paper.
%
CB acknowledges financial support from the Deutsche Forschungsgemeinschaft.

\end{document}